\documentclass[12pt]{article}
\usepackage{latexsym}
\usepackage{amsmath,amsfonts}
\usepackage{times}
\allowdisplaybreaks[4]

\begin{document}

\begin{flushright}
FIAN-TD-2017-26  \qquad \ \ \ \ \ \ \ \ \ \ \\
arXiv: 1711.00272v1 [hep-th] \\
\end{flushright}

\vspace{1cm}

\begin{center}

{\Large \bf Secondary fields and partial wave expansion. Self consistency conditions in a conformal model.
}

\vspace{2.5cm}

V.N. Zaikin%
\footnote{ E-mail: zaykin@lpi.ru
}

\vspace{1cm}

{\it I.E Tamm Theory Department, P.N. Lebedev Physical Institute\\
Leninsky prospekt 53, 119991, Moscow, Russia }

\vspace{3.5cm}

{\bf Abstract}

\end{center}

     A nontrivial conformally invariant model is obtained via generalization the method of obtaining conformally invariant models in $2D$ Euclidean space to the Euclidean space with dimension \\ $D>2$. This method was  previously developed by E.S. Fradkin and M.Ya. Palchik (see [7] and reference therein).
     The partial wave expansion of a four-point function $\langle j_{\mu}(x_{1})j_{\nu}(x_{2})\varphi(x_{3})\varphi(x_{4})\rangle$ containing two conserved vector fields $j_{\mu}$ and two scalars $\varphi$ of dimension $ d $ in a $ D $ -dimensional Euclidean space is considered. The requirement of the absence the vector operator of the dimension $ d + 1 $ in this expansion allows us to find the relationship between all the coupling constants  in such a model.

\newpage
\renewcommand{\thefootnote}{\arabic{footnote}}
\setcounter{footnote}{0}

\section{  Introduction  }

The relationship between operator product expansion in conformal field theory and short distance limits for various conformal invariant correlation functions was established many years ago [1-8]. Of particular interest is the study of four-point functions of primary (fundamental) fields. In recent years, significant progress has been achieved  in analyzing such functions [9-12].

In the works of E. S. Fradkin and M.Ya. Palchik [4] it was shown that if the field $P_{s}^{\delta}$ of the spin $s$ and dimension $\delta$ contributes to the operator product expansion of any fields $A (x)B (y)$, then the density of the partial wave expansion $\rho(l,\sigma)$ of a multipoint function of the type $\langle A(x)B(y)...\rangle$ has a first-order pole at the point $l=\delta, \sigma=s$ and vice versa, the presence of a first-order pole in the density of the partial wave expansion means that in the operator product expansion $A (x)B (y)$ there is a descendant field $P^{\delta}_{s}$. Also it was shown that in any conformal invariant theory of a scalar or spinor field with the scalar dimension $d$ there exists a set of tensor or spin-tensor fields:
\begin{equation}\label{}
 P_{s}(x)\equiv P_{\mu_{1}\mu_{2}...\mu_{s}}^{d+s}(x), \nonumber
\end{equation}
generated by energy-momentum tensor or conserved current. Each of the fields $P_{s}(x)$ is a symmetric traceless field of the spin $s$ and scalar dimension $d+s$.These fields together with all their derivatives emerge in the operator product expansions:
\begin{equation}\label{}
  T_{\mu\nu}(x_{1})\varphi(x_{2}),\quad j_{\mu}(x_{1})\varphi(x_{2}),\quad j_{\mu}(x_{1})\psi(x_{2})\nonumber,
\end{equation}
where $T_{\mu\nu}$ is an energy-momentum tensor, $j_{\mu}$ is a conserved current, $\psi(y)$ and $\varphi(y)$ are the spinor and scalar fields.
It was suggested to consider the fields $P_{s}(x)$ for $D>2$ as analogs of the secondary fields and the states $P_{s}\mid 0>$ as analogs of null-vector of the 2-dimensional conformal theories. In two-dimensional space a condition
 \begin{equation}\label{}
   P_{s}\mid 0>=0 \nonumber
 \end{equation}
 allows one to find exact solutions of various conformal models. In the works of E.S. Fradkin and M.Ya. Palchik  it was proposed to extend the requirement $P_{s}\mid 0>=0$ to the case $D>2$ and it was suggested [5], that such a generalization should correspond to yet unknown representation of $D$-dimensional analog of Virasoro algebra.
In 2-dimensional Euclidean space, the operator products $j_{\mu}(x)\psi(y)$ and $T_{\mu\nu}(x)\varphi(y)$ and the partial wave expansion of the four-point functions \\
$ <j_{\mu}(x_{1})j_{\nu}(x_{2})\psi(y_{1})\overline{\psi}(y_{2})> $ \, and \,
 $ <T_{\mu\nu}(x_{1})T_{\rho\sigma}(x_{2})\varphi(y_{1})\varphi(y_{2})> $ \, were studied in the detail [6]. It has been shown , that the densities of the partial wave expansions of these functions $\rho_{j}(l,\sigma)$ and $\rho_{T}(l,\sigma)$ have poles of the first order at $l=d+s,\sigma=s$. Demanding the residue at the pole $l = d + 1/2, \sigma=1/2$ for the function $\rho_{j}(l,\sigma)$ or the residue at the pole $l = d + 2, \sigma=2$ for the function $\rho_{T}(l,\sigma)$ be zero, which corresponds to the absence the state $P_{s}^{d+s}\mid 0>$ in the dynamical sector of the model, gives algebraic equations on the dimensions of all fields included in these theories as well as coupling constants and differential equations for higher correlation functions [7,13]. The results coincided with those obtained earlier by other methods for the Ising, Thirring, and others models.

  In this paper, an attempt is made to generalize the above method of constructing conformal invariant models to the D-dimensional Euclidean space. For simplicity we consider the 4-point function $<j_{\mu}(x_{1})j_{\nu}(x_{2})\varphi(y_{1})\varphi(y_{2})>$ for the two conserved vector fields $j_{\mu}(x)$ of  dimension $d_{j} = D-1$ and two charged scalar fields $\varphi(y)$ of  dimension $d$.

\section{Basis functions of partial wave expansion   }
We investigate the four-point function, containing two conserved currents  $j_{\mu}(x)$ and scalar fields $\varphi^d(x)$ of the scalar dimension $d_{j}=D-1$ and $d$, respectively.
\begin{eqnarray}
\label{1}
  G_{\mu\nu}(x_{1},x_{2}|x_{3},x_{4})=<j_{\mu}(x_{1})j_{\nu}(x_{2})\varphi(x_{3})\varphi(x_{4})>
\end{eqnarray}

Partial wave expansion of such function can be represented in the form:
\begin{eqnarray}\label{2}
&& G_{\mu\nu}(x_{1},x_{2}|x_{3},x_{4})=\nonumber\\
&&\sum_{i,j=1}^2\int d\sigma\rho_{ij}(\sigma)C_{i\,\{\mu_{s}\};\mu}^{ld_{j}d}(x_{a};x_{1}|x_{3})
\left[D_{\left\{\mu_{s}\right\}\left\{\nu_{s}\right\}}^l(x_{ab})\right]^{-1}C_{j\,\{\nu_{s}\};\nu}^{ld_{j}d}(x_{b};x_{2}|x_{4})\,d^{\small D}x_{a}\,d^{\small D}x_{b}
\end{eqnarray}
 Here the following notations and definitions are used:
\begin{eqnarray}
&&\int\,d\sigma=\frac{1}{2\pi i}\sum_{s}\int_{h-i\infty}^{h+i\infty}\mu(l,s)\,dl\,,\quad h=\frac{D}{2};\\
&&\mu(l,s)=2^{s-2}(2\pi)^{-h}\frac{\Gamma(h+s)}{\Gamma(s+1)}n(l,s)n(\tilde l,s), \quad \tilde{l}=D-l \nonumber\\
&& n(l,s)=\frac{\Gamma(l+s)\Gamma(D-l-1)}{\Gamma(h-l)\Gamma(D-l+s-1)}\nonumber
\end{eqnarray}

$\left[D_{\{\mu_{s}\}\{\nu_{s}\}}^l(x_{ab})\right]^{-1}$ \mbox{is the inverse propagator of the traceless symmetric field $P_{\left\{\mu_{s}\right\}}^l(x)$}, of the spin $s$ and scalar dimension $l$,\\
 \begin{eqnarray}\label{}
&&D_{\{\mu_{s}\}\{\nu_{s}\}}^l(x_{a}-x_{b})=
\frac{1}{(2\pi)^h}2^{l}n(l,s)\{g_{\mu_{1}\nu_{1}}(x_{ab})\cdots g_{\mu_{s}\nu_{s}}(x_{ab})\mbox {- traces}\}(x_{ab}^2)^{-l},\\
&&\left[D_{\{\mu_{s}\}\{\nu_{s}\}}^l(x_{a}-x_{b})\right]^{-1}=D_{\{\mu_{s}\}\{\nu_{s}\}}^{\tilde l}(x_{a}-x_{b}),\quad \tilde{l}=D-l
\end{eqnarray}
 \qquad\qquad $g_{\mu\nu}(x)=\frac{1}{x^2}\left(\delta_{\mu\nu}-2\frac{x_{\mu}x_{\nu}}{x^2}\right),$ \qquad $x_{ab}=x_{a}-x_{b}$ \\

 $C_{i\,\{\mu_{s}\};\mu}^{ld_{j}d}(x_{1};x_{2}|x_{3}), i=1,2$ are three-point functions which transform under conformal\\
  transformations as \quad $<j_{\mu}(x_{1})\varphi(x_{2})P_{\left\{\mu_{s}\right\}}^l(x_{3})>$.

It is known that the partial wave expansion (2) is related with the operator product expansion. Namely,
the first-order poles of the spectral densities $ \rho_{ij}(l, s) $ correspond to the  descendant  fields
in the operator product expansion $ j_{\mu} (x) \varphi (0) $ and vice versa, to each descendant field arising in the operator
product expansion, there corresponds a first-order pole of the function $ \rho_{ij}(l, s) $. The residue at this pole is proportional to the contribution of the descendant field to the operator product expansion.

To find the explicit form of $\rho_{ij}(l,s)$ it is necessary at first to choose the basis functions\\
$C_{i\,\{\mu_{s}\};\mu}^{ld_{j}d}(x_{1};x_{2}|x_{3})$ over which the decomposition is carried out. As mentioned above,\\
functions $C_{i\,\{\mu_{s}\};\mu}^{ld_{j}d}(x_{1};x_{2}|x_{3})$ have the same conformal properties as  3-point
function \\ $<j_{\mu}(x_{1})\varphi(x_{2})P_{\left\{\mu_{s}\right\}}^l(x_{3})>$.
There exist just two linearly independent conformal structures that transform
 as $<j_{\mu}(x_{1})\varphi(x_{2})P_{\left\{\mu_{s}\right\}}^l(x_{3})>$:
\begin{eqnarray}
&&A_{1\,\{\mu_{s}\};\mu}^{ld_{j}d}(x_{1};x_{2}|x_{3})=\lambda_{\mu}^{x_{2}}(x_{1},x_{3})\left[\lambda_{\mu_{1}\mu_{2}\ldots\mu_{s}}^{x_{1}}(x_{2},x_{3})-traces\right]
\Delta(x_{1},x_{2},x_{3}),\\
&&A_{2\,\{\mu_{s}\};\mu}^{ld_{j}d}(x_{1};x_{2}|x_{3})=\frac{1}{x_{12}^{2}}\left[\sum_{k=1}^s g_{\mu\mu_{k}}(x_{12})\lambda_{\mu_{1}\ldots\hat \mu_{k}\ldots\mu_{s}}^{x_{1}}(x_{2},x_{3})-traces\right]\Delta(x_{1},x_{2},x_{3}),
\end{eqnarray}
\begin{eqnarray}
&&\lambda_{\mu}^{x_{2}}(x_{1},x_{3})=\frac{(x_{12})_{\mu}}{x_{12}^2}-\frac{(x_{32})_{\mu}}{x_{32}^2},\\
&&\lambda_{\mu_{1}\mu_{2}\ldots\mu_{s}}^{x_{1}}(x_{2},x_{3})=\lambda_{\mu_{1}}^{x_{1}}(x_{2},x_{3})\cdots\lambda_{\mu_{s}}^{x_{1}}(x_{2},x_{3}),\\
&&\Delta(x_{1},x_{2},x_{3})=(x_{12}^2)^{-\frac{l-s+d_{j}-1-d}{2}}(x_{13}^2)^{-\frac{l-s-d_{j}+1+d}{2}}(x_{23}^2)^{-\frac{-l+s+d_{j}-1+d}{2}},
\end{eqnarray}
$\hat{\mu_{k}}$  -- means omitting the index.

The basis functions are some two linearly independent combinations of conformal structures {}\\
 $A_{i} (i=1,2)$ .

If $d_{j}=D-1$, than due to the presence of the factor $(x_{12}^2)^{-\frac{l-s+d_{j}-1-d}{2}}$ both structures $A_{1}$ and $A_{2}$ have the first order poles at the points $l=d+s$, the residues in which are related by the relation [7]:

\begin{equation}\label{}
   res\left[(D-2)A_{1}+A_{2}\right]\big|_{l=d+s}=0.
\end{equation}
The following linearly independent structures were chosen as the basic functions:
\begin{equation}\label{}
   C_{long}=N_{long}[A_{1}+p_{long}(l,d)A_{2}], \qquad   p_{long}(l,d)=\frac{1}{D-2+l-d-s},
\end{equation}
\begin{equation}\label{}
  C_{tr}=N_{tr}[A_{1}+p_{tr}(l,d)A_{2}], \qquad   p_{tr}(l,d)=\frac{l-d}{s(D-2+l-d-s)}.
\end{equation}
With this choice, both functions $C_{long}$ and $C_{tr}$ are regular at $l=d+s$, function $C_{long}$  is local and $C_{tr}$ is transversal (for the details see [7]):
\begin{equation}\label{}
    \partial_{\mu}^{x_{2}}C_{tr\,\left\{\mu_{s}\right\}\,\mu}^{ld_{j}d}(x_{1};x_{2}|x_{3})\big|_{d_{j}=D-1}=0
\end{equation}
In addition, it turns out that so selected basic functions with an appropriate choice of \\  normalizations $N_{long}$ and $N_{tr}$ satisfy for tensor field the "amputation" condition :
\begin{equation}
\int\left[D_{\mu_{s}\nu_{s}}^l(x_{1}-x_{a})\right]^{-1}C_{long,tr\{\nu_{s}\};\mu}^{ld_{j}d}(x_{a};x_{2}|x_{3})\,d^{\small D}x_{a}=
C_{long,tr\{\mu_{s}\};\mu}^{\tilde l d_{j}d}(x_{1};x_{2}|x_{3}),
\end{equation}

 \begin{eqnarray}
 && N_{long}=n_{long}(l)\left\{2^{\frac{2l-d}{2}}\frac{(D-2+l-d-s)}{(2D-2-l-d-s)}
 {\frac{\Gamma(\frac{l-d+s+D-2}{2})\Gamma(\frac{l+d+s-D+2}{2})}{\Gamma(\frac{2D-2-l-d+s}{2})
 \Gamma(\frac{-l+2+s+d}{2})}}\right\}^{\frac{1}{2}},\nonumber\\
 && N_{tr}=n_{tr}(l)\left\{2^{\frac{2l-d}{2}}\frac{(l-1)(D-2-l+d+s)}{(D-l-1)(2D-2-l-d-s)}
 {\frac{\Gamma(\frac{l-d+s+D-2}{2})\Gamma(\frac{l+d+s-D+2}{2})}{\Gamma(\frac{2D-2-l-d+s}{2})\Gamma(\frac{-l+2+s+d}{2})}}\right\}^{\frac{1}{2}}.\nonumber
 \end{eqnarray}
 $n_{long}(l)$ and $n_{tr}(l)$ are arbitrary functions satisfying conditions $n_{long, tr}(l)=n_{long, tr}(\tilde l)$.\\

Functions $C_{long\{\mu_{s}\};\mu}^{l_{1}\tilde d_{j}\tilde d}$ and  $C_{tr\{\mu_{s}\};\mu}^{l_{1}\tilde d_{j}\tilde d}$ orthonormal to $C_{long\{\mu_{s}\};\mu}^{ld_{j}d}$ and $C_{tr\{\mu_{s}\};\mu}^{ld_{j}d}$ also have been found:

\begin{equation}\label{}
  C_{long\{\mu_{s}\};\mu}^{l_{1}\tilde d_{j}\tilde d}\big|_{\tilde{d}_{j}=1}=N^{\ast}_{long}(l,d)\left[A_{1\{\mu_{s}\};\mu}^{l_{1}\tilde d_{j}\tilde d}+p^{\ast}_{long}(l,d)A_{2\{\mu_{s}\};\mu}^{l_{1}\tilde d_{j}\tilde d}\right]\Big|_{\tilde{d}_{j}=1},
\end{equation}
\begin{equation}
  p^{\ast}_{long}(l,d)=\frac{1}{l-D+d-s},\nonumber
\end{equation}
and
\begin{equation}
C_{tr\{\mu_{s}\};\mu}^{l_{1}\tilde d_{j}\tilde d}\big|_{\tilde{d}_{j}=1}=N^{\ast}_{tr}(l,d)\left[A_{1\{\mu_{s}\};\mu}^{l_{1}\tilde d_{j}\tilde d}+p^{\ast}_{tr}(l,d)A_{2\{\mu_{s}\};\mu}^{l_{1}\tilde d_{j}\tilde d}\right]\Big|_{\tilde{d}_{j}=1},
\end{equation}

\begin{equation}
  p^{\ast}_{tr}(l,d)=\frac{l-2D+d+2}{s(-l+3D-4-d+s)}.\nonumber
\end{equation}
Orthonormality conditions for these functions are:
\begin{eqnarray}
&&\int C_{long,tr\{\mu_{s_{1}}\};\mu}^{l_{1}\tilde d_{j}\tilde d}(x_{1};x_{a}|x_{b})
C_{long,tr\{\mu_{s_{2}}\};\nu}^{l_{2}d_{j}d}(x_{2};x_{a}|x_{b})d^{\small D}x_{a}d^{\small D}x_{b}= \nonumber\\
&&=\frac{1}{2}\left[D_{\{\mu_{s_{1}}\}\{\nu_{s_{1}}\}}^{l_{1}}(x_{1}-x_{2})\delta_{\sigma_{1}\sigma_{2}}+\hat I_{\{\mu_{s_{1}}\}\{\nu_{s_{1}}\}}\delta(x_{1}-x_{2})\delta_{\sigma_{1}\tilde \sigma_{2}}\right],\\
&&\int C_{long,tr\{\mu_{s_{1}}\};\mu}^{l_{1}\tilde d_{j}\tilde d}(x_{1};x_{a}|x_{b})
C_{tr,long\{\mu_{s_{2}}\};\nu}^{l_{2}d_{j}d}(x_{2};x_{a}|x_{b})dx_{a}dx_{b}=0,
\end{eqnarray}

$\delta_{\sigma_{1}\sigma_{2}}=\delta(l_{1}-l_{2})\delta_{s_{1}s_{2}}$,\quad $\hat{I}_{\{\mu_{s}\}\{\nu{s}\}}$ is unit operator:
$\hat{I}_{\{\mu_{s}\}\{\nu_{s}\}} F_{\{\mu_{s}\}}=F_{\{\mu_{s}\}}.$

In addition it turns out that these functions   $C_{long,tr\{\mu_{s}\};\mu}^{l\tilde d_{j}\tilde d}$ with the appropriate choice \\
 of normalizations $N_{long,tr}^{\ast}(l,d)$ also satisfy for tensor field the "amputation" condition:
\begin{equation}
\int\left[D_{\mu_{s}\nu_{s}}^l(x_{1}-x_{a})\right]^{-1}C_{long,tr\{\nu_{s}\};\mu}^{l\tilde d_{j}\tilde d}(x_{a};x_{2}|x_{3})\,d^{\small D}x_{a}=
C_{long,tr\{\mu_{s}\};\mu}^{\tilde l \tilde d_{j}\tilde d}(x_{1};x_{2}|x_{3}) ,\quad  \tilde l=D-l
\end{equation}
 Condition (20) fixes the normalizations $N_{long,tr}^{\ast}(l,d)$ up to invariant functions
 \begin{eqnarray}
 && N_{long}^{\ast}(l,d)=n_{long}^{\ast}(l)\left\{2^{\frac{2l-d}{2}}\frac{(d-s+l-D)}{(d-s-l)}
 {\frac{\Gamma(\frac{l-d+s+D}{2})\Gamma(\frac{l+d+s-D}{2})}{\Gamma(\frac{2D-l-d+s}{2})
 \Gamma(\frac{-l+s+d}{2})}}\right\}^{\frac{1}{2}}\nonumber\\
 && N_{tr}^{\ast}(l,d)=n_{tr}^{\ast}(l)\left\{2^{\frac{2l-d}{2}}\frac{(l-1)(-3D+4+l+d-s)}{(D-l-1)(-2D+4-l+d-s)}
 {\frac{\Gamma(\frac{l+D-d+s}{2})\Gamma(\frac{l-D+d+s}{2})}
 {\Gamma(\frac{2D-l-d+s}{2})\Gamma(\frac{-l+s+d}{2})}}\right\}^{\frac{1}{2}}\nonumber\\
 && n_{long,tr}^{\ast}(l)=n_{long,tr}^{\ast}(\tilde{l}) \nonumber
 \end{eqnarray}

\section{  Densities of the partial wave expansion and fixing the model }
The partial wave expansion (2) can be written in the basis (12), (13) as follows:
\begin{eqnarray}\label{}
&&G_{\mu\nu}(x_{1},x_{2}|x_{3},x_{4})=\nonumber\\
&&\sum_{\sigma}\,d\sigma\rho_{long,long}(\sigma)C_{long\,\{\mu_{s}\};\mu}^{ld_{j}d}(x_{a};x_{1}|x_{3})
\left[D_{\mu_{s}\nu_{s}}^l(x_{a}-x_{b})\right]^{-1}C_{long\,\{\nu_{s}\};\nu}^{ld_{j}d}(x_{b};x_{2}|x_{4})\,dx_{a}\,dx_{b}+\nonumber\\
&&\sum_{\sigma}\,d\sigma\rho_{long,tr}(\sigma)C_{long\,\{\mu_{s}\};\mu}^{ld_{j}d}(x_{a};x_{1}|x_{3})
\left[D_{\mu_{s}\nu_{s}}^l(x_{a}-x_{b})\right]^{-1}C_{tr\,\{\nu_{s}\};\nu}^{ld_{j}d}(x_{b};x_{2}|x_{4})\,dx_{a}\,dx_{b}+\nonumber\\
&&\sum_{\sigma}\,d\sigma\rho_{tr,long}(\sigma)C_{tr\,\{\mu_{s}\};\mu}^{ld_{j}d}(x_{a};x_{1}|x_{3})
\left[D_{\mu_{s}\nu_{s}}^l(x_{a}-x_{b})\right]^{-1}C_{long\,\{\nu_{s}\};\nu}^{ld_{j}d}(x_{b};x_{2}|x_{4})\,dx_{a}\,dx_{b}+\nonumber\\
&&\sum_{\sigma}\,d\sigma\rho_{tr,tr}(\sigma)C_{tr\,\{\mu_{s}\};\mu}^{ld_{j}d}(x_{a};x_{1}|x_{3})
\left[D_{\mu_{s}\nu_{s}}^l(x_{a}-x_{b})\right]^{-1}C_{tr\,\{\nu_{s}\};\nu}^{ld_{j}d}(x_{b};x_{2}|x_{4})\,dx_{a}\,dx_{b}.
\end{eqnarray}
Using the orthogonality conditions (18),(19), we obtain the following equations on the densities of the partial wave expansion (2):
\begin{eqnarray}\label{}
&&\int C_{long\{\mu_{s}\};\mu}^{l\tilde d_{j}\tilde d}(x_{1};x_{a}|x_{b})G_{\mu\nu}(x_{a},x_{2}|x_{b},x_{3})dx_{a}dx_{b}=\nonumber\\
&&\rho_{long,long}(l)C_{long\{\mu_{s}\};\nu}^{ld_{j}d}(x_{1};x_{2}|x_{3})+
 \rho_{long,tr}(l)C_{tr\{\mu_{s}\};\nu}^{ld_{j}d}(x_{1};x_{2}|x_{3})
\end{eqnarray}
and similarly
\begin{eqnarray}\label{}
&&\int C_{tr\{\mu_{s}\};\mu}^{l\tilde d_{j}\tilde d}(x_{1};x_{a}|x_{b})G_{\mu\nu}(x_{a},x_{2}|x_{b},x_{3})dx_{a}dx_{b}=\nonumber\\
&&\rho_{tr,long}(l)C_{long\{\mu_{s}\};\nu}^{ld_{j}d}(x_{1};x_{2}|x_{3})+
 \rho_{tr,tr}(l)C_{tr\{\mu_{s}\};\nu}^{ld_{j}d}(x_{1};x_{2}|x_{3}).
\end{eqnarray}
Recall that the dimension of the conserved current is $ d_ {j} = D-1 $. It turns out that in this case the function
$C_{long\{\mu_{s}\};\mu}^{l\tilde d_{j}\tilde d}(x_{1};x_{a}|x_{b})$ can be represented as a total derivative:
\begin{equation}\label{}
  C_{long\{\mu_{s}\};\mu}^{l\tilde d_{j}\tilde d}(x_{1};x_{a}|x_{b})\big|_{d_{j}=D-1}=N^{\ast}(l,d) \partial_{\mu}^{x_{a}} <P_{\mu_{s}}^l(x_{1})j^{(0)}(x_{a})\varphi^{\tilde d}(x_{b})>,
  \end{equation}
  where
  \begin{eqnarray}\label{}
&& <P_{\{\mu_{s}\}}^l(x_{1})j^{(0)}(x_{a})\varphi^{\tilde d}(x_{b})>=\nonumber\\
&&\left[\lambda_{\mu_{1}\mu_{2}\ldots\mu_{s}}^{x_{1}}(x_{a}x_{b})-traces\right](x_{1a}^2)^{-\frac{l-s-D+d}{2}}
(x_{1b}^2)^{-\frac{l-s+D-d}{2}}(x_{ab}^2)^{-\frac{-l+s+D-d}{2}}.
\end{eqnarray}
    Substituting (24) into the left-hand side of equation (22) and integrating by parts, we find:
   \begin{eqnarray}
   &&\int C_{long\{\mu_{s}\};\mu}^{l\tilde d_{j}\tilde d}(x_{1};x_{a}|x_{b})G_{\mu\nu}(x_{a},x_{2}|x_{b},x_{3})dx_{a}dx_{b}=\nonumber\\
   &&-N^{\ast}(l,d)\int \left <P_{\{\mu_{s}\}}^l(x_{1})j^{(0)}(x_{a})\tilde \varphi(x_{b})\right>\partial_{\mu}^{x_{a}}G_{\mu\nu}(x_{a},x_{2}|x_{b},x_{3})dx_{a}dx_{b}
   \end{eqnarray}

   For the further analysis it is necessary to make an assumption concerning the divergency of the function $G_{\mu\nu}(x_{a},x_{2}|x_{b},x_{3})$.The following Ansatz consistent with the requirement of conformal invariance is taken:
   \begin{eqnarray}
   && \partial_{\mu}^{x_{a}}G_{\mu\nu}(x_{a},x_{2}|x_{b},x_{3})=-{\large\bf{e}}[\delta (x_{a}-x_{b})-\delta (x_{a}-x_{3})]\left<j_{\nu}(x_{2})\varphi(x_{b})\varphi(x_{3})\right> +\qquad\nonumber\\
   && {\bf{Q}}\partial_{\nu}^{x_{a}}\delta(x_{a}-x_{2})P(x_{2};x_{b},x_{3}){}+{}{\bf{G}}\partial_{\mu}^{x_{a}}\delta(x_{a}-x_{2})T_{\mu\nu}(x_{2};x_{b},x_{3})+\nonumber\\ &&{\bf{C}}\partial_{\mu}^{x_{a}}D_{\mu\nu}^{d_{j}}(x_{a}-x_{2})D^d(x_{b}-x_{3})
    \end{eqnarray}
    where
     \begin{eqnarray}
     &&\left<j_{\nu}(x_{2})\varphi(x_{b})\varphi(x_{3})\right>=-\textbf{e}\frac{\Gamma(\frac{D}{2})}{2\pi ^{\frac{D}{2}}}{}\lambda_{\nu}^{x_{2}}(x_{b}x_{3})(x_{b2}^2x_{32}^2)
      ^{-\frac{D-2}{2}}(x_{b3}^2)^{-\frac{2d-D+2}{2}}\\
     &&P(x_{2};x_{b},x_{3}){}={}(x_{2b}^2x_{23}^2)^{-\frac{D-2}{2}}(x_{b3}^2)^{-\frac{2d-D}{2}}\\
     && T_{\mu\nu}(x_{2};x_{b},x_{3}){}={}\Large\left[\lambda_{\mu}^{x_{2}}(x_{b}x_{3})\lambda_{\nu}^{x_{2}}(x_{b}x_{3})-\frac{1}{D}\delta_{\mu\nu}\frac{x_{b3}^2}{x_{b2}x_{32}}\Large\right]
     (x_{2b}^2x_{23}^2)^{-\frac{D-4}{2}}(x_{b3}^2)^{-\frac{2d-D+4}{2}}
     \end{eqnarray}
     $D_{\mu\nu}^{d_{j}}(x_{a}-x_{2})$ and $G^d(x_{b}-x_{3})$ are some functions that transform under conformal
     transformations in the same manner as propagators of vector and scalar fields of dimensions $d_ {j}$ and $d$, respectively
     \begin{eqnarray}
     &&D_{\mu\nu}^{d_{j}}(x_{a}-x_{2}){}={}N(d_{j})g_{\mu\nu}(x_{a2})(x_{a2}^2)^{-d_{j}}\\
     &&D^d(x_{b}-x_{3}){}={}N(d)(x_{b3}^2)^{-d}
     \end{eqnarray}
     The normalizations of these functions are chosen from conditions:
     \begin{eqnarray}
 &&\int D_{\mu_{1}\nu}^{d_{j}}(x_{1}-x)[D_{\mu_{2}\nu}^{d_{j}}(x-x_{2})]^{-1}d^{D}x=\delta_{\mu_{1}\mu_{2}}\delta(x_{1}-x_{2})\\
 &&\int D^d(x_{1}-x)[D^d(x-x_{2})]^{-1}d^{D}x=\delta(x_{1}-x_{2})
\end{eqnarray}

  The question of the uniqueness of the representation (27) is open and needs further investigation.
  The expression (27) is one of the conditions fixing the model. It turns out that for such a choice of the function $\partial_{\mu}^{x_{a}}G_{\mu\nu}(x_{a},x_{2}|x_{b},x_{3})$ the integral (22) of each of the
  terms on the right-hand side of (26)\,can be found in an explicit form. Similar integrals can be found in [7],[14].

  The only exception is the integral of the contribution proportional to $ \textbf{C} $.
   To find this contribution, it is necessary to regularize the integrand in the original integral (22). \\
   Let:
  \begin{eqnarray}
   && I_{\left\{\mu_{s}\right\};\nu}^{ld_{j}d}(x_{1};x_{3}|x_{2})\big|_{d_{j}=D-1}\equiv \nonumber\\
   &&\int\left <P_{\left\{\mu_{s}\right\}}^l(x_{1})j^{(0)}(x_{a})^{\tilde{d}}\varphi(x_{b})\right>\partial_{\mu}^{x_{a}}D_{\mu\nu}^{d_{j}}(x_{a}-x_{2})D^d(x_{b}-x_{3})d^Dx_{a}d^Dx_{b}\big|_{d_{j}=D-1}=\nonumber\\
   &&-\lim_{d_{j}\to D-1}\int\left[C_{\left\{\mu_{s}\right\}long}^{l\tilde{d_{j}}\tilde{d}}(x_{1};x_{a}|x_{b})\right]^{reg}D_{\mu\nu}^{d_{j}}(x_{a}-x_{2})D^d(x_{b}-x_{3})d^Dx_{a}d^Dx_{b}
  \end{eqnarray}
 where
  \begin{eqnarray}
  &&\left[C_{long\,\{\mu_{s}\};\,\mu}^{l\tilde{d_{j}}\tilde{d}}(x_{1};x_{a}|x_{b})\right]^{reg}=\alpha_{long}(l,d)A_{1\mu_{s}:\mu}^{ld_{j}d}(x_{1};x_{a}|x_{b})+A_{2\mu_{s}:\mu}^{ld_{j}d}(x_{1};x_{a}|x_{b})\\
  &&\alpha_{long}(l,d)=l-s+d-D+(d_{j}-D+1)\textbf{r}(l,d)
\end{eqnarray}
and the regularization parameter $\textbf{r}(l,d)$ is an arbitrary function of the variables $l$ and $d$ . The need to introduce this parameter will be explained below.
 The result of integrating each of the terms in (35) contains first-order pole in the variable $d_{j}$  at $d_{j}=D-1$. However, with the chosen value $\alpha_{long}$, these poles cancel each other and in the resulting expression one can take the limit $d_{j}=D-1$. After simple, but rather tedious calculations, we obtain:
 \begin{eqnarray}
 &&\lim_{d_{j}\to D-1}\int\left[C_{long\,\{\mu_{s}\};\,\mu}^{l\tilde{d_{j}}\tilde{d}}(x_{1};x_{a}|x_{b})\right]^{reg}D_{\mu\nu}^{d_{j}}(x_{a}-x_{2})D^d(x_{b}-x_{3})d^Dx_{a}d^Dx_{b}=\nonumber\\
 &&\rho^{[\bf{C}]}_{long,long}(l,d)\frac{1}{N_{long}(l,d)}C_{long\,\{\mu_{s}\};\nu}^{l\,D-1\,d}(x_{1};x_{3}|x_{2})+\nonumber\\
 &&\rho^{[\bf{C}]}_{long,tr}(l,d)\frac{1}{N_{tr}(l,d)}C_{tr\,\{\mu_{s}\};\nu}^{l\,D-1\,d}(x_{1};x_{3}|x_{2}),
\end{eqnarray}
where
\begin{eqnarray}
&&\rho^{[\bf{C}]}_{long,long}(l,d)=\frac{1}{l-d-s}\rho_{0}(l,d),\\
&&\rho^{[\bf{C}]}_{long,tr}(l,d)=\frac{1}{l-d-s}\rho_{1}(l,d),\\
&&\rho_{0}(l,d)=2^{d+1}\frac{(D-2+l-d-s)}{(D-2+l-d+s)}\frac{\Gamma(\frac{D-l+d+s}{2})\Gamma(\frac{l+d+s}{2})}{\Gamma(\frac{l-d+s}{2})\Gamma(\frac{D-l-d+s}{2})},\nonumber\\
&&\rho_{1}(l,d)=s(D-2)\nonumber\\
&&\frac{\left[-4(l-1)(D-2d)+(l-d-s)(D-2+l-d+s)\textbf{r}(l,d)\right]}{(l-d+s)(D-l-d+s)(l+d+s-2)(D-2+l-d-s)}\rho_{0}(l,d).
\end{eqnarray}
Let us note that the regularization parameter $\textbf{r}(l,d)$ appears only in the expression for $\rho^{[\bf{C}]}_{long,tr}(l,d)$.
It turns out that for the obvious identity following from (38)

\begin{eqnarray}
&& \hspace{-0.9cm} \lim_{d_{j}\to D-1}\int\!\! d^D\!x_{a}d^D\!x_{b} \! \left[D_{\left\{\mu_s\right\}\left\{\nu_s\right\}}^l(x_1-x)\right]^{-1}\!\!
\left[C_{long\,\{\mu_{s}\};\,\mu}^{l\tilde{d_{j}} \tilde{d}}(x;x_{a}|x_{b})\right]^{reg}\!\! D_{\mu\nu}^{d_{j}}(x_{a}-x_{2})D^d(x_{b}-x_{3})
\nonumber\\
&& \hspace{-0.9cm} = \rho^{[\bf{C}]}_{long,long}(l,d) \frac{1}{N_{long}(l,d)} \int\! d^D\!x \left[D_{\left\{\mu_s\right\} \left\{\nu_s\right\}}^l(x_1-x)\right]^{-1}C_{long\,\{\mu_{s}\};\,\nu}^{lD-1d}(x;x_{3}|x_{2})
\nonumber\\
&& \hspace{-0.9cm} + \,\, \rho^{[\bf{C}]}_{long,tr}(l,d)\frac{1}{N_{tr}(l,d)} \int\! d^D\!x \left[D_{\left\{\mu_s\right\} \left\{\nu_s\right\}}^l(x_1-x)\right]^{-1}C_{tr\,\{\mu_{s}\};\,\nu}^{lD-1d}(x;x_{3}|x_{2})
\end{eqnarray}
to hold, the regularization function $\textbf{r}(l,d)$ must be the solution to some functional equation. An analogous equality occurs upon integrating both parts of (38) with the function
$\left[D^{d}(x-x_{3})\right]^{-1}$.

For example, the condition for the fulfillment of (42) gives the equation:
\begin{equation}
  \frac{(l-1)(2D-l-d-s-2)}{(D-l-1)(D+l-d-s-2)}\omega(\tilde{l},d) = \omega(l,d),
\end{equation}
where

\begin{equation}\label{}
  \omega(l,d)=
  \frac{(D-2)}{(D-2+l-d-s)}[(l-d-s)(D-2+l-d+s)(1+\textbf{r}(l,d))-4(l-1)(D-2d)].
\end{equation}
A similar equation arises from the requirement of self-consistency of both sides of equality (42) with "amputation" over the scalar field.

The general solution to these self-consistency equations can be represented as:
\begin{equation}\label{}
  \mathbf{r}(l,d)=-1+\frac{(l-1)(D+s-l-d)(D-s-l-d)(l-d+s)}{(D-2+l-d+s)}\mathbf{r_{0}}(l,d),
\end{equation}
 where $\mathbf{r_{0}}(l,d)$ is an arbitrary function satisfying the following conditions:
\begin{equation}\label{}
\mathbf{r_{0}}(l,d)=\mathbf{r_{0}}(\tilde{l},d)=\mathbf{r_{0}}(l,\tilde{d}).
\end{equation}
Note that  $\mathbf{r}(l,d)$ can not be equal to zero for any choice of the  $\mathbf{r_{0}}(l,d)$

The contributions of the remaining terms on the right-hand side of (27) to the integral (22) can be represented as:
\begin{eqnarray}
  &&\rho_{long,long}^{\Sigma}=\frac{1}{(l-d-s)^2} \Phi_{long}^{(2)}(l,d)+\frac{1}{(l-d-s)} \Phi_{long}^{(1)}(d) + \{...\}, \\
  &&\rho_{long,tr}^{\Sigma}=\frac{1}{(l-d-s)^2} \Phi_{tr}^{(2)}(l,d)+\frac{1}{(l-d-s)} \Phi_{tr}^{(1)}(d) + \{...\},
\end{eqnarray}
where $\{...\}$ denotes terms having no pole at $l=d+s$
\begin{eqnarray}
&&\Phi_{long}^{(2)}(l,d)=s(d+s-1){\it{k}}_{long}(l,d)\nonumber\\
&&\left[(D-2)(2d-D+2){\bf{Q}}+\frac{2}{D}(-sD(s+d-1)+(2-dD-3D+2d+D^2)){\bf{G}}\right],\\
&&\Phi_{tr}^{(2)}(l,d)=s(d+s-1){\it{k}}_{tr}(l,d)\nonumber\\
&&\left[(D-2)(2d-D+2){\bf{Q}}+\frac{2}{D}(-sD(s+d-1)+(2-dD-3D+2d+D^2)){\bf{G}}\right],
\end{eqnarray}
where
\begin{eqnarray}
&&{\it{k}}_{long}(l,d)=\frac{(D-2+l-d-s)}{(D-2+l-d+s)}{\it{k}}_{0}(l,d),\\
&&{\it{k}}_{tr}(l,d)=\frac{(D-2-l+d+s)}{(D-2+l-d+s)}{\it{k}}_{0}(l,d),\\
&&{\it{k}}_{0}(l,d)=\pi^{\frac{D}{2}}\frac{\Gamma(\frac{2+d+s-l}{2})\Gamma(\frac{d+s+l-D}{2})\Gamma(D-d)}{\Gamma(\frac{D+l-d+s}{2})\Gamma(\frac{2D-l+s-d}{2})\Gamma(d-\frac{D}{2}+2 )},
\end{eqnarray}
\begin{eqnarray}
&&\Phi_{long}^{(1)}(l,d)=(-1)^{s+1}\frac{(D-2+l-d-s)}{(D-2+l-d+s)}{\bf{e}}+\\
&&s(D-2s-2d){\it{k}}_{long}(l,d)\left[(2d-D+2){\bf{Q}}+\frac{2}{D}(sD-1-d+dD){\bf{G}}\right],\\
&&\Phi_{tr}^{(1)}(l,d)=s(l-1)(D-2s-2d){\it{k}}_{tr}(l,d){\bf{G}}.
\end{eqnarray}
The total contribution of all terms of the expression (27) to the integral (22) is equal to:
\begin{eqnarray}
&&\rho_{long,long}(l,d)=\rho_{long,long}^{\left[\Sigma\right]}(l,d)+\rho_{,long,long}^{\left[C\right]}(l,d) \\
&&\rho_{long,tr}(l,d)=\rho_{long,tr}^{\left[\Sigma\right]}(l,d)+\rho_{long,tr}^{\left[C\right]}(l,d)
\end{eqnarray}

Let us require now the spectral densities (58) and (59) to be free of the second order poles in the variable $l$ at  $l=d+s$. This means that the following equations must be fulfilled:
\begin{equation}\label{}
\Phi_{long}^{(2)}(l,d)=0 \qquad \Phi_{tr}^{(2)}(l,d)=0
\end{equation}
This is possible only if
\begin{equation}\label{}
  \left[(D-2)(2d-D+2){\bf{Q}}+\frac{2}{D}\left(-sD(s+d-1)+(2-dD-3D+2d+D^2)\right){\bf{G}}\right]=0.
\end{equation}
Further fixing of the model according to the procedure proposed earlier is to require the absence of certain descendant field in the operator product expansion $j_{\mu}(x)\varphi(0)$.
This requirement means that the residue of the spectral functions (58) and (59) at the pole corresponding to this descendant field must vanish. Let us consider the simplest example
when there is no descendant vector field of dimension $l=d+1$ in the expansion under consideration. In this case, equating to zero the residues at poles of the first order of the spectral
densities $\rho_{long,long}(l,d)$  and $\rho_{long,tr}(l,d)$, and taking into account condition (61), we obtain two more algebraic equations:
\begin{eqnarray}
 &&(D-2){\large\bf{e}}-8d\frac{(D-1)}{D}\frac{\pi^\frac{D}{2}}{\Gamma(\frac{D}{2})}{\bf{G}}+2^{d+1}\frac{\Gamma(d+1)}{\Gamma(\frac{D}{2}-d)}(D-2)\Gamma(\frac{D}{2}){\bf{C}}=0,\\
 &&4d\frac{(D-2)}{D}\frac{\pi^\frac{D}{2}}{\Gamma(\frac{D}{2})}{\bf{G}}+2^{d+1}\frac{\Gamma(d+1)}{\Gamma(\frac{D}{2}-d)}(D-2)\Gamma(\frac{D}{2}){\bf{C}}=0
 \end{eqnarray}
These equations together with (61) allow us to find the relationships between the free parameters of the model:
\begin{eqnarray}
&&{\bf{G}}=\frac{D(D-2)}{4d(3D-4)}\frac{\Gamma(\frac{D}{2})}{\pi^{\frac{D}{2}}}\,{\large\bf{e}},\\
&&{\bf{C}}=-2^{-d-1}\frac{(D-2)}{(3D-4)}\frac{\Gamma(\frac{D}{2}-d)}{\Gamma(d+1)}\frac{1}{\Gamma(\frac{D}{2})}\,{\large\bf{e}},\\
&&{\bf{Q}}=\frac{(D-1)}{2d(3D-4)}\frac{\Gamma(\frac{D}{2})}{\pi^{\frac{D}{2}}}\,{\large\bf{e}}.
\end{eqnarray}
\section{  Conclusion }
In the $D$-dimensional Euclidean space, a conformally invariant model is constructed that includes a charged scalar field and a conserved vector current.
The method of obtaining such a model is based on a generalization of some results of the method of constructing conformally invariant models in 2-dimensional Euclidean space, proposed earlier in the works of Palchik and Fradkin [7]. It can be assumed that in an $D$-dimensional Euclidean space there is an analogue of an infinite parametrical Virasoro algebra [15].  The condition that the state $P_{s}\mid 0>$ and all its descendants be equal to zero in this case can point out at the irreducibility of the representation of such an algebra. As in the two-dimensional case, this is possible only for certain values of both the dimensions of the fundamental fields and other parameters. Correlation functions of  these fields must satisfy partial differential equations of finite order.
As the first step of this program the relationships between the constants that fix the model are found. Later it is expected to obtain closed differential equations for multi-point correlation functions, consider models involving spinor fields, and also models with non-Abelian fields. Of particular interest are the models, including energy-momentum tensor.
\section*{Acknowledgements}
The author is thankful to Mikhail A. Vasiliev and Ruslan R. Metsaev for very useful discussions and comments on the draft of the paper.

\end{document}